\documentstyle[11pt,epsfig]{article}
\begin{document}
\begin{titlepage}

\begin{flushright} 
  \bf{CMS NOTE 1999/057}
\end{flushright} 
\begin{center}
  {\Large \bf Muon Track Matching}\\
\end{center}

\begin{center}
    A. Benvenuti\\
INFN, Bologna, Italy\\
 D. Denegri\\
DAPNIA, Saclay, France\\
 V. Genchev, P. Vankov\\
INRNE, Sofia, Bulgaria\\
 A. Khanov, N. Stepanov\\
ITEP, Moscow, Russia\\
\end{center}

\begin{center}
   (November 1, 1999)
\end{center}

  \begin{abstract}
 For most physical processes the tracks observed in the muon stations 
must be matched with the corresponding tracks in the inner tracker,
the external muon system providing muon identification and triggering
but the tracker points giving the precise momentum measurement at lower
momenta. For high momenta the momentum resolution is much improved by the
interconnection of inner and outer measurements. The matching of outer and 
inner measurements is more delicate in case of muons embedded in jets. A 
study of the matching procedure was carried out using samples of $b\bar{b}$ 
jets at high P$_{t}$, requiring $b\bar{b}$ $\rightarrow$ $\mu$ decays.
  \end{abstract} 
  
\end{titlepage}

\section{Motivation}

 Identification and measurement of muons in the inner tracker using its 
high spatial resolution and excellent momentum measurement in the 4T field 
is a very important and demanding task, particularly for muons in jets. In 
the muon system the occupancy is low and pattern recognition is relatively 
easy, provided the particle reaches the muon system at all, which is the 
case for the barrel is P$_{t}\geq$ 4 GeV. For non-isolated muons the main 
difficulty is to match the track observed in the muon stations with the 
correct candidate in the inner tracker in case of the high track density 
inside a jet. These situations arise for example in muon tagging of $b$-jets,
in $b$ $\rightarrow$ $J/\psi$($\rightarrow$ $\mu$ $\mu$) + $X$ cascades in
$b$ and $top$ physics, in SUSY($\bar{q}$, $\bar{g}$) physics and most 
dramatically in very high E$_{t}$ (TeV scale) jets such as heavy flavour 
($b$, $c$) QCD jets giving rise to "muon bundles" within jets. 

 The matching strategy is based on the use of the full parameter space, 
i.e. on both the reconstructed momentum and the position in a given 
plane for the two tracks, one in the muon system and the other 
in the inner tracker. The quantities monitored are the efficiency of 
matching (number of muons matched over number of muons reconstructed),
the muon matching impurity (1 - number of muons matched over number of
tracks matched) and generation of "ghost" and "fake" tracks. 

\section{CMS model and data sample}

 The results discussed here are obtained using a detailed GEANT description of the 
complete CMS detector of TDR design. In all simulations the effects of the 
4T field are fully included. 

 We have performed a GEANT simulation of the response to $b\bar{b}$ jets 
with transverse momenta P$_{t}^{jet}$ of 50, 100 and 200 GeV 
(which corresponds to the P$_{t}$ range of $b$-jets in $t\bar{t}$ 
production) and at least one muon inside each jet with P$_{t}^{\mu}$ 
$\geq$ 5 GeV in the rapidity region -1.1 $\leq$ $\eta$ $\leq$ 1.1. 
1000 events were generated and reconstructed in the barrel part at each 
energy point. 

 The results are obtained using the standard CMS reconstruction 
procedure (data card 'MUON' 2 and 'TRAK' 3) existing in the CMSIM version 115. 

 Pattern recognition in the muon stations is done by building track 
segments from the reconstructed hits separately in the (r,$\phi$) and 
(r,z) planes; (r$\phi$,rz) track segments in the same muon station and 
in the same or contiguous (r,$\phi$) sectors are considered to form a pair 
used to perform the Kalman filter. This is done starting from the outermost 
stations and proceeding towards the inner stations with a constrained fit 
to the impact point (IP) with default values of $\sigma_{x,y}^{IP}$ = 
15 $\mu m$ and $\sigma_{z}^{IP}$ = 5.3 cm. Since the decay mean transverse 
length (impact parameter) of the
$b$-quarks is about 500 $\mu m$ another reconstruction was made using a 
constrained fit to the impact point with 
$\sigma_{x,y}^{IP}$ = 500 $\mu m$ and $\sigma_{z}^{IP}$ = 5.3 cm.

The  Global Track Finder (GTF) was used for pattern recognition in the CMS 
inner tracker. The GTF algorithm adopted for the CMS tracker must process a 
large number of hits per event: typically 5.10$^{3}$ hits at low luminosity 
and ten times more at high luminosity. To overcome the severe 
combinatorial problems, the concept of road preselection is used in the
first stage of the algorithm. In the second stage a Kalman filter is
used to carry out final hit selection and track fitting.

\section{Track matching}

 Given two track segments, one in the muon system and the other in 
the inner tracker, one can extrapolate both segments to a common plane 
or point, check the difference in momentum and position and
select the best combination among the candidates. 
The correct combination is known from the simulation procedure and it is 
therefore possible to calculate the efficiency of matching.
We use as a matching criterion the minimal distance computed from $\eta$, 
$\phi$ and/or P$_{t}^{rec}$ of the two track candidates
at the outer plane (TRCI - CMSIM convention) of the inner tracker and at 
the impact point. Association of reconstructed 
tracks to GEANT ones is done on the hit basis. The reconstructed 
track is associated to a given GEANT particle number if more than half of the 
hits assigned to the track are from that particular GEANT track. 
If there is no such dominating track, the reconstructed track is called
a "ghost". If more than one reconstructed track is matched to a given 
GEANT number it is called a "fake".

 The $\eta$ dependence of the reconstruction efficiency obtained for 
the barrel muon station, shown on fig. 1a) is rather uniform except for the 
region between the central and next-to-central rings ($\mid\eta\mid$ $\leq$ 0.2) and 
for $\mid\eta\mid$ $\geq$ 0.9. To avoid the influence of the $\eta$ 
dependence of the reconstruction efficiency on the matching results the 
tracks with $\mid\eta\mid$ $\geq$ 0.9 are removed. The P$_{t}$ reconstruction 
efficiency (fig. 1b)) is more than 95\% for the tracks with P$_{t}^{\mu}$ $\geq$ 
10 GeV. In fig. 1c) is given the "type" of reconstructed tracks in the muon 
stations (solid line) for which must be found the best candidate between
the tracks reconstructed in the tracker (error bars).
The comparison between reconstructed and generated variables: P$_{t}$, 
$\eta$ and $\phi$ is given in fig. 1d) - fig. 1i) both for the barrel muon 
stations and for the inner tracker, for P$_{t}^{jet}$ = 100 GeV.
We find a P$_{t}$ muon resolution of about 10\% for the muon stations alone, 
and of $\leq$ 1\% for the inner tracker, in good agreement with the values 
given in the Muon and Tracker TDR's$^{/1, 2/}$. 

To estimate the matching quality a $\chi^{2}$ is formed from the 
differences between reconstructed $\eta$, $\phi$ and P$_{t}$ in
the two detectors for all reconstructed track combinations at the outer tracker 
plane and at the impact point: 

\begin{equation}
 \chi^{2} = \sqrt{\frac{(\eta_{tr}^{rec} - \eta_{ms}^{rec})^{2}}{\sigma_{\eta}^{2}} 
+ \frac{(\phi_{tr}^{rec} - \phi_{ms}^{rec})^{2}}{\sigma_{\phi}^{2}} + \frac{(Pt_{tr}^{rec} - Pt_{ms}^{rec})^{2}}{\sigma_{Pt}^{2}}}
\end{equation}

 Including in the $\chi^{2}$, equation (1), the contribution from
P$_{t}^{rec}$ in the muon stations and the tracker improves the 
selection of the matching tracks mainly for the high transverse momenta. 
A minimal $\chi^{2}$ of the matching tracks has been used as a matching 
criterion. 

 The P$_{t}$ dependence of the matching efficiency is shown on fig. 2
for P$_{t}^{jet}$ = 50, 100 and 200 GeV at the outer plane TRCI of the inner 
tracker and at the interaction point with the two assumptions for the 
constrained fit to the impact point with 
$\sigma_{x,y}$ = 15 $\mu m$ and $\sigma_{x,y}$ = 500 $\mu m$.
The degradation of the matching efficiency at higher P$_{t}^{jet}$ is
is due to the increase of charged particle multiplicity inside
jets which makes more difficult the correct association of the tracks observed 
in the muon stations with the track candidates in the inner tracker. 
The matching efficiency shows less variation when a constrained fit at 
the impact point is performed using $\sigma_{x,y}$ = 500 $\mu m$. 

 Results obtained from the matching of muons inside the $b\bar{b}$ jets at 
P$_{t}^{jet}$ of 50, 100 and 200 GeV are summarized in Table 1. The "type" 
of the matched tracks indicates a higher purity at the impact point
and at lower jet momenta. For the higher multiplicity and more collimated jets at P$_{t}^{jet}$ = 
200 GeV, the impurity in matching degrades to a $\sim$ 7\% level. Rejecting 
tracks with an increasing momentum threshold $\leq$ P$_{t}^{cut}$, a high matching efficiency and 
purity of track matching can be obtained for high P$_{t}^{jet}$, but at 
the expense of a substantial reduction in statistics. The number of 
"ghost" tracks generated at reconstruction in the muon stations is negligible.
The number of "fake" tracks shows some tendency to increase with jets 
momenta, and is at level of 1\% - 2\%.

 After performing the matching procedure we can assign to the tracks 
observed in the muon station the precise measurements from the tracker. 
The distributions of the differences between reconstructed and generated 
values of P$_{t}$, $\eta$ and $\phi$ at the impact point
for the matched tracks at P$_{t}^{jet}$ = 100 GeV are given in fig. 3. 
These distributions are practically the same 
as in fig. 2, where we compared reconstructed and generated variables for 
the inner tracker, confirming the good quality of the matching criterion.
The bottom row in fig. 3 gives the "type" of the tracks reconstructed in the 
muon stations and the "type" of matched tracks at the impact point.

 The quality of track matching can be measured by the value of the "pull":

\begin{equation}
   P = \frac{z_{tr}P_{tr} - z_{ms}P_{ms}}{P_{ms}^{2}/\sqrt{cov(1/p.1/p)}}, 
\end{equation}

where:

z$_{tr}$, z$_{ms}$ is the track charge in the tracker and the muon station;

P$_{tr}$, P$_{ms}$ is the reconstructed momentum in tracker and muon station;

cov(1/p.1/p) is the covariance matrix.

 A Gaussian fit to the pull shown at the bottom of fig. 3 gives $\sigma$ 
$\sim$ 1.0 indicating a good accuracy for the selection and matching of 
the tracks.

\section{Summary and conclusion} 
 
 The main results concerning the matching of muon tracks in ($b\bar{b}$) 
jets between the muon stations and inner tracker are:

Using a simple minimal distance criterion between $\eta$, $\phi$ and 
P$_{t}$ of independent track measurements at the impact point
one can obtain a high matching efficiency with a contamination of not more 
than 4\% for P$_{t}^{jet}$ of 50 GeV and 6\% for 100 GeV. For the more 
collimated jets with P$_{t}^{jet}$ = 200 GeV, the impurity in matching is 
at a $\sim$ 7\% level.

The matching efficiency shows less variation when a constrained fit at 
the impact point is performed with $\sigma_{x,y}$ = 500 $\mu m$. 
To further improve the matching efficiency and impurity a more sophisticated 
software than currently available must be used, for example a constrain
fir to the secondary $b\bar{b}$ jets vertices obtained from the inner
tracker can be performed. 
 
Rejecting tracks with an increasing momentum cut-off $\leq$ P$_{t}^{cut}$, a high purity of track 
matching can be obtained for hard jets, but at the expense of a 
substantial reduction of the statistics.

~\\~\\

\begin{table}[hbtp]
\caption{Track matching results.}
    \label{tab:match1}
	\begin{center}
\footnotesize{
\tiny{
\begin{tabular}{|c|c|c|c|c|c|c|c|} \hline
 \multicolumn{2}{|c|}{P$_{t}^{jet}$ = 50 GeV} & \multicolumn{2}{|c|}{TRCI} & \multicolumn{2}{|c|}{IP $\sigma_{x,y}$ = 15 $\mu m$} & \multicolumn{2}{|c|}{IP $\sigma_{x,y}$ = 500 $\mu m$} \\ \hline \hline
 P$_{t}^{cut}$ (GeV) & loss of stat. (\%) & efficiency (\%) & impurity (\%) & efficiency (\%) & impurity (\%) & efficiency (\%) & impurity (\%) \\ \hline
 5 & 0 & 96.42 & 3.64 $\pm$ 0.47 & 96.22 & 4.08 $\pm$ 0.50 & 96.83 & 3.23 $\pm$ 0.45 \\ \hline
 6 & 2.68 & 96.85 & 2.89 $\pm$ 0.43 & 96.98 & 3.02 $\pm$ 0.48 & 97.06 & 2.81 $\pm$ 0.43 \\ \hline
 7 & 7.07 & 97.53 & 2.00 $\pm$ 0.37 & 97.73 & 1.86 $\pm$ 0.33 & 97.63 & 1.89 $\pm$ 0.36 \\ \hline
 8 & 11.98 & 97.97 & 1.60 $\pm$ 0.34 & 98.26 & 1.38 $\pm$ 0.30 & 98.16 & 1.11 $\pm$ 0.29 \\ \hline
 9 & 17.97 & 98.13 & 1.41 $\pm$ 0.33 & 98.52 & 1.09 $\pm$ 0.24 & 98.43 & 0.79 $\pm$ 0.25 \\ \hline
 10 & 24.79 & 98.47 & 0.85 $\pm$ 0.27 & 98.73 & 0.68 $\pm$ 0.23 & 98.46 & 0.60 $\pm$ 0.23 \\ \hline
 \multicolumn{2}{|c|}{"ghost" tracks(\%)} & \multicolumn{4}{|c|}{0.13 $\pm$ 0.09} & \multicolumn{2}{|c|}{0.06 $\pm$ 0.06} \\ \hline 
 \multicolumn{2}{|c|}{"fake" tracks(\%)} & \multicolumn{4}{|c|}{1.47 $\pm$ 0.30} & \multicolumn{2}{|c|}{1.45 $\pm$ 0.39} \\ \hline \hline
 \multicolumn{2}{|c|}{P$_{t}^{jet}$ = 100 GeV} & \multicolumn{2}{|c|}{TRCI}  & \multicolumn{2}{|c|}{IP $\sigma_{x,y}$ = 15 $\mu m$} & \multicolumn{2}{|c|}{IP $\sigma_{x,y}$ = 500 $\mu m$} \\ \hline \hline
 P$_{t}^{cut}$ (GeV) & loss of stat. (\%) & efficiency (\%) & impurity (\%) & efficiency (\%) & impurity (\%) & efficiency (\%) & impurity (\%) \\ \hline
 5 & 0 & 93.57 & 6.55 $\pm$ 0.61 & 94.40 & 6.24 $\pm$ 0.62 & 94.87 & 5.89 $\pm$ 0.59 \\ \hline
 6 & 2.22 & 93.93 & 6.13 $\pm$ 0.60 & 95.18 & 5.24 $\pm$ 0.56 & 95.61 & 5.05 $\pm$ 0.55 \\ \hline
 7 & 4.32 & 94.38 & 5.68 $\pm$ 0.59 & 95.66 & 4.71 $\pm$ 0.58 & 96.43 & 4.13 $\pm$ 0.51 \\ \hline
 8 & 6.49 & 94.78 & 5.28 $\pm$ 0.57 & 96.16 & 4.16 $\pm$ 0.54 & 96.94 & 3.44 $\pm$ 0.47 \\ \hline
 9 & 9.57 & 95.15 & 4.85 $\pm$ 0.56 & 96.17 & 3.96 $\pm$ 0.56 & 97.12 & 3.15 $\pm$ 0.46 \\ \hline
 10 & 13.03 & 95.66 & 4.27 $\pm$ 0.54 & 96.73 & 3.27 $\pm$ 0.48 & 97.30 & 2.84 $\pm$ 0.44 \\ \hline
 15 & 26.03 & 97.39 & 2.36 $\pm$ 0.44 & 98.06 & 1.77 $\pm$ 0.33 & 98.83 & 1.17 $\pm$ 0.31 \\ \hline
 \multicolumn{2}{|c|}{"ghost" tracks(\%)} & \multicolumn{4}{|c|}{0.08 $\pm$ 0.07} & \multicolumn{2}{|c|}{0.10 $\pm$ 0.08} \\ \hline 
 \multicolumn{2}{|c|}{"fake" tracks(\%)} & \multicolumn{4}{|c|}{1.78 $\pm$ 0.41} & \multicolumn{2}{|c|}{1.36 $\pm$ 0.38} \\ \hline \hline
 \multicolumn{2}{|c|}{P$_{t}^{jet}$ = 200 GeV} & \multicolumn{2}{|c|}{TRCI}  & \multicolumn{2}{|c|}{IP $\sigma_{x,y}$ = 15 $\mu m$} & \multicolumn{2}{|c|}{IP $\sigma_{x,y}$ = 500 $\mu m$} \\ \hline \hline
 P$_{t}^{cut}$ (GeV) & loss of stat. (\%) & efficiency (\%) & impurity (\%) & efficiency (\%) & impurity (\%) & efficiency (\%) & impurity (\%) \\ \hline
 5 & 0 & 90.17 & 9.99 $\pm$ 0.74 & 92.55 & 8.07 $\pm$ 0.67 & 94.12 & 6.90 $\pm$ 0.62 \\ \hline
 6 & 0.60 & 90.60 & 9.51 $\pm$ 0.72 & 93.00 & 7.51 $\pm$ 0.69 & 94.33 & 6.59 $\pm$ 0.61 \\ \hline
 7 & 1.62 & 91.18 & 8.94 $\pm$ 0.71 & 93.42 & 6.98 $\pm$ 0.66 & 94.70 & 6.12 $\pm$ 0.60 \\ \hline
 8 & 3.25 & 91.72 & 8.40 $\pm$ 0.69 & 93.69 & 6.72 $\pm$ 0.67 & 95.00 & 5.48 $\pm$ 0.57 \\ \hline
 9 & 5.05 & 91.94 & 8.17 $\pm$ 0.69 & 94.15 & 6.14 $\pm$ 0.68 & 95.24 & 5.07 $\pm$ 0.56 \\ \hline
 10 & 6.80 & 92.24 & 7.88 $\pm$ 0.68 & 94.56 & 5.75 $\pm$ 0.52 & 95.21 & 5.04 $\pm$ 0.56 \\ \hline
 15 & 17.63 & 93.69 & 6.38 $\pm$ 0.65 & 95.98 & 4.23 $\pm$ 0.53 & 96.44 & 3.70 $\pm$ 0.51 \\ \hline
 20 & 23.95 & 94.86 & 5.07 $\pm$ 0.62 & 96.75 & 3.25 $\pm$ 0.41 & 97.50 & 2.65 $\pm$ 0.46 \\ \hline
 \multicolumn{2}{|c|}{"ghost" tracks(\%)} & \multicolumn{4}{|c|}{0.18 $\pm$ 0.10} & \multicolumn{2}{|c|}{0.18 $\pm$ 0.10} \\ \hline 
 \multicolumn{2}{|c|}{"fake" tracks(\%)} & \multicolumn{4}{|c|}{1.99 $\pm$ 0.34} & \multicolumn{2}{|c|}{2.11 $\pm$ 0.35} \\ \hline \hline
 \end{tabular}}}
 \end{center}
   \end{table}

\newpage

\begin{figure}[hbtp]
  \begin{center}
    \resizebox{10cm}{!}{\includegraphics{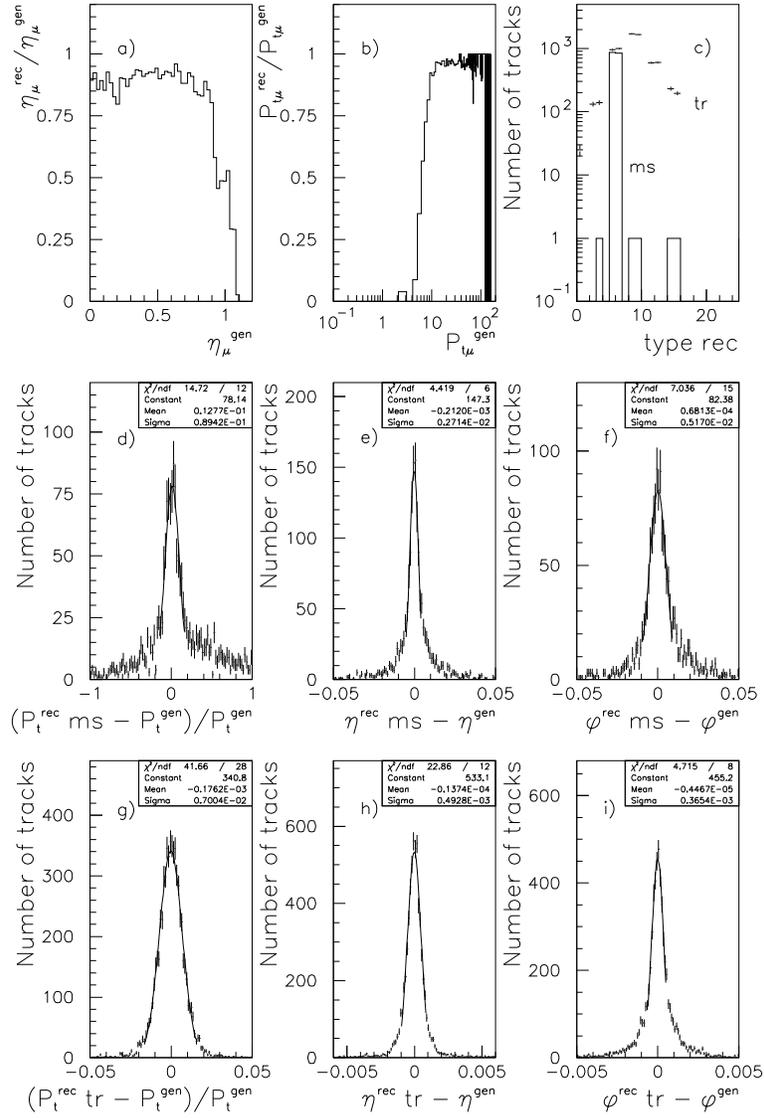}}
\caption{$\eta$ (a) and P$_{t}$ (b) dependence of the reconstructed 
efficiency, "type" of reconstructed tracks (c) in the muon stations (solid
line) and in the tracker (error bars) and comparison 
between reconstructed and generated variables P$_{t}$, $\eta$ and $\phi$ 
in the barrel muon station (d - f) and in the inner tracker (g - i) for 
P$_{t}^{jet}$ = 100 GeV.}
\label{fig:1}
  \end{center}
\end{figure}
 
\newpage

\begin{figure}[hbtp]
  \begin{center}
    \resizebox{10cm}{!}{\includegraphics{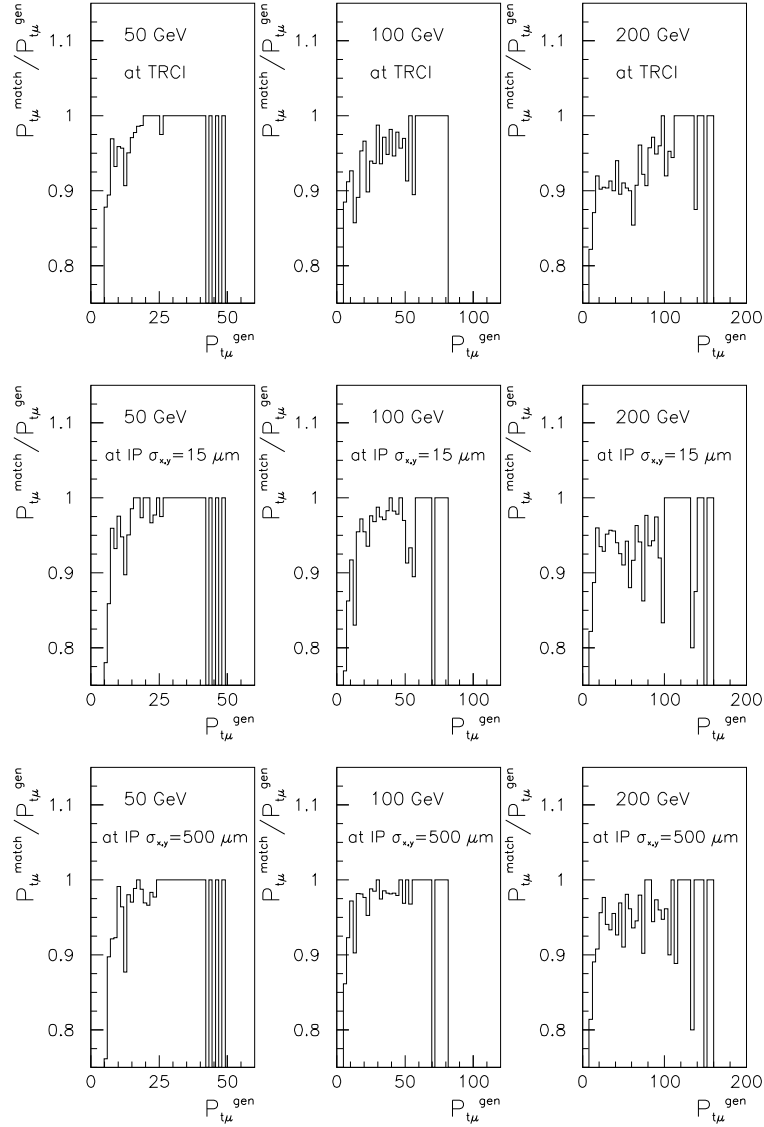}}
\caption{P$_{t}$ dependence of the matching efficiency for P$_{t}^{jet}$ = 50,
100 and 200 GeV at the outer plane TRCI of the inner tracker 
and at the impact point with two assumptions for the 
constrained fit to the impact point with 
$\sigma_{x,y}$ = 15 $\mu m$ and $\sigma_{x,y}$ = 500 $\mu m$ respectively.}
\label{fig:2}
  \end{center}
\end{figure}
 
\newpage

\begin{figure}[hbtp]
  \begin{center}
    \resizebox{10cm}{!}{\includegraphics{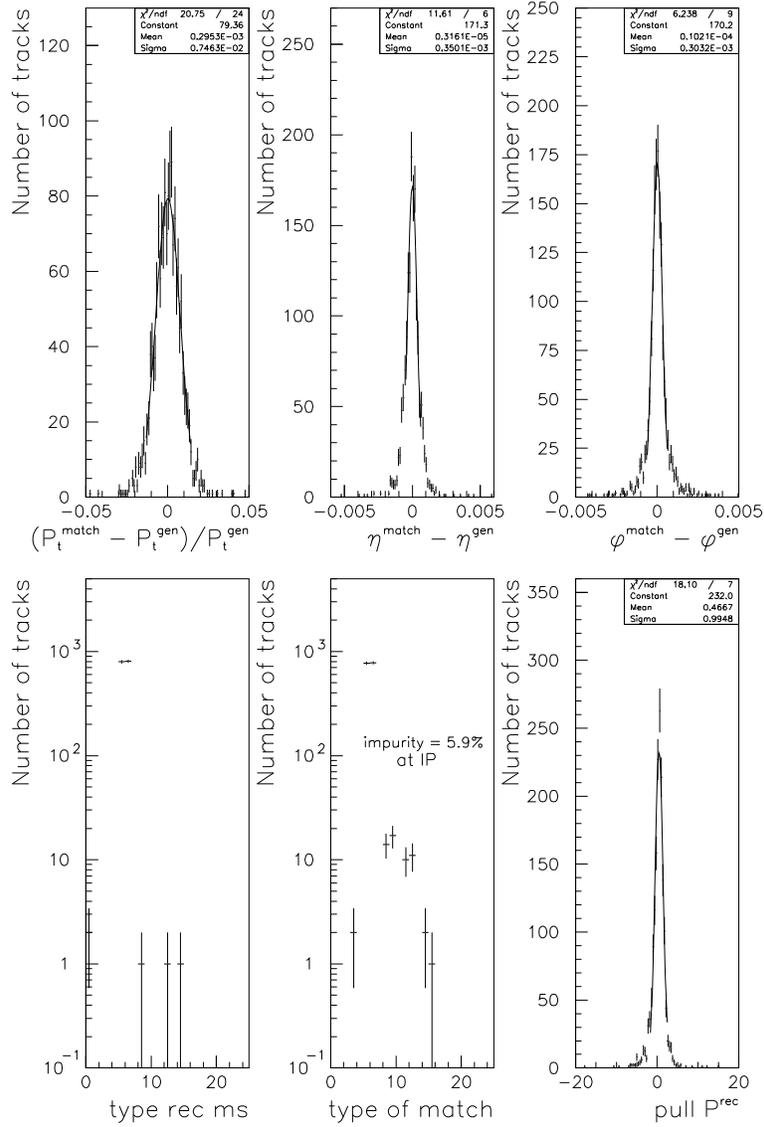}}
\caption{Comparison of  reconstructed and generated values of
P$_{t}$, $\eta$ and $\phi$, "type" and  "pull" for matched tracks at
impact point for P$_{t}^{jet}$ = 100 GeV.}
\label{fig:3}
  \end{center}
\end{figure}
 
\end{document}